\documentclass[aps,superscriptaddress,showpacs,nofootinbib]{revtex4}
\usepackage{graphics}

\input{epsf}

\newcommand{\be}{\begin{equation}}
\newcommand{\ee}{\end{equation}}
\newcommand{\bea}{\begin{eqnarray}}
\newcommand{\eea}{\end{eqnarray}}
\newcommand{\lessa}{\stackrel{<}{\sim}}

\begin{document}

\preprint{UG-FT-177/04}

\preprint{CAFPE-47/04}

\title{Determining the regimes of cold and warm inflation in the SUSY
  hybrid model} 
\author{Mar Bastero-Gil}
\email{mbg@ugr.es}
\affiliation{Departamento de F\'{\i}sica Te\'orica y del Cosmos,
  Universidad de Granada, Granada-18071, Spain}

\author{Arjun Berera}
\email{ab@ph.ed.ac.uk}
\affiliation{ School of Physics, University of
Edinburgh, Edinburgh, EH9 3JZ, United Kingdom}

\begin{abstract}
 
The SUSY hybrid inflation model is found to dissipate radiation
during the inflationary period.  Analysis is made of parameter regimes
in which these dissipative effects are significant.  The scalar
spectral index, its running, and the tensor-scalar ratio
are computed in the entire parameter range of the model.
A clear prediction for strong dissipative warm inflation is found 
for $n_S-1 \simeq 0.98$ and a low tensor-scalar ratio much below
$10^{-6}$.  The strong dissipative warm inflation regime also
is found to have no $\eta$-problem and the field amplitude much
below the Planck scale.  As will be discussed,
this has important theoretical
implications in permitting a much wider variety of SUGRA
extensions to the basic model.
 
\medskip                                   

\noindent
keywords: cosmology, inflation
\end{abstract}
                                                                                

\maketitle

\section{Introduction}

The success of the inflationary paradigm has motivated in recent times
more serious efforts in building realistic particle physics
models that incorporate cosmology \cite{kk,shafi,bck}.  
The objective of this sort
of model building is to account for various cosmological features,
central being inflation, but also leptogenesis, dark matter etc...
to constrain the high energy properties of the model,
and such that in the low energy regime the model reduces to the
Standard Model.  Supersymmetry (SUSY) plays a central role here,
since aside from its attractive features for particle physics,
it also allows stabilizing very flat scalar potentials, which
are essential in inflation models due to density perturbation constraints.
In this respect, a widely studied SUSY model of inflation has
been the hybrid model,
\be
W= \kappa S ( \Phi_1 \Phi_2 - \mu^2) \, ,
\label{superpot}
\ee
where $\Phi_1$, $\Phi_2$ are a pair of charged fields\footnote{One
could also consider the fields $\Phi_{1,2}$ being gauge singlets. However,
in this case the gravitino constraint on the reheating temperature
translates into $\kappa$ being at most of the order of
$10^{-5}$ \cite{mohapatra}. That constraint is avoided when instead
$\Phi_{1,2}$ are non-singlets.} under some gauge
group $G$, and $S$ is the singlet which plays the role of the
inflaton.

An important feature about the model Eq. (\ref{superpot})
and its various embeddings into more realistic particle
physics models \cite{jkls,shafi} is that 
the inflaton field generally interacts
with other fields, with coupling strengths that can
be fairly large.  Even though the nonzero vacuum energy
necessary to drive inflation will break SUSY, this underlying symmetry
can still protect the very flat inflaton potential 
from radiative corrections arising from these perturbatively
large couplings \cite{br4,hm}.  It has been observed in recent
works \cite{br4,br} that the effect of interactions of the inflaton
with other fields
does not simply affect
the local contributions to the inflaton effective potential,
but also induces temporally nonlocal terms in the inflaton
evolution equation, that
in the moderate to
large perturbative regime
yield sizable dissipative effects.
Although SUSY cancels the large local quantum
effects, for the dynamical problem the nonlocal
quantum effects can not be canceled by SUSY.
These dissipative effects in general can lead
to warm inflationary regimes \cite{wi}.   Thus the conclusion of
the works \cite{br4,br,hm} is that in general, models in which
the inflaton has interactions with other fields with
moderate to strong perturbative coupling, inflation divides
into two different dynamical regimes, cold 
\cite{oldi,ni,ci} and warm \cite{wi}.
This finding is very important, since these two types of
inflationary dynamics are qualitatively much different.
Thus one should expect different observable signatures
in the two cases, as well as other theoretical differences
in the treatment of inflation.

The purpose of this paper is to apply these recent findings about
dissipative dynamics during inflation to the SUSY hybrid model
Eq. (\ref{superpot}) and to common extensions of this model.
In particular, two models will be studied in this paper,
Eq. (\ref{superpot}) and this model with a 
matter field $\Delta$ ${\bar \Delta}$
coupled to it as
\be
W= \kappa S ( \Phi_1 \Phi_2 - \mu^2) + g \Phi_2 \Delta \bar \Delta \,.
\label{superpot2}
\ee
The above is a toy model representing an example of how
the basic hybrid model Eq. (\ref{superpot}) is embedded within
a more complete particle physics model, in this case through
the $\Delta$ fields.

In this paper we will study inflation for both models
Eqs. (\ref{superpot}) and (\ref{superpot2}).
We will show that in the above models both cold and warm inflation
exist and we will determine the parameter regime for
them.  This will then explicitly verify the conclusions
from the recent papers on dissipation \cite{br4,br}, that
showed both types of inflationary dynamics could exist.
In both inflationary regimes, we will calculate the scalar spectral
index $n_S-1$, and its running $dn_S/d\ln k$.
With this information, we will then identify the qualitative
and quantitative differences arising from the warm versus
cold regimes.
 
We emphasize here that the main objective of this paper is
to determine in an explicit and well-known model
how these two dynamically very different inflationary
regimes emerge.  This result departs radically
from current wisdom, in which it is tacitly always assumed
that all regimes in any model are cold inflationary.
In light of our result, many other particle physics model
building issues emerge in the warm inflationary regimes
that we identify.  Although we make some effort
in this paper to discuss these issues, it is not the
purpose here to dwell on them.  In cold inflation,
these issues have been the subject of many years of study, and
likewise a complete understanding of similar issues for
warm inflationary regimes will take focused effort in
future work.

The paper is organized as follows.  Sect. \ref{sect2} 
reviews the effective potential for the SUSY hybrid model,
as well as some of the basic results based on this model for
cold inflation.  Sect. \ref{sect3} reviews the main results of
dissipative dynamics during inflation based on \cite{br4,br}.
Included here are the main formulas for density perturbations
and the scalar spectral index in the presence of dissipation
and a thermal bath.  Subsects. \ref{sect3a} and \ref{sect3b}
compute the effect that dissipation has on inflaton evolution
in the two models Eqs. (\ref{superpot}) and (\ref{superpot2})
respectively. The outcome of the analysis in these
two subsections is a graph that divides the parameter
space of the model into cold and warm inflationary regimes,
and associated predictions for the scalar spectral
index, its running and the tensor-scalar ratio.  Sect. \ref{sect4}
addresses some model building issues that emerge in the newly
found warm inflationary regimes, such as the gravitino abundances and
the constraint on the reheating temperature. 
Finally Sec. V states our conclusions.

\section{Susy Hybrid Model}
\label{sect2}

We briefly review first some well known results about
standard supersymmetric hybrid inflation
\cite{copeland,dvali,lazarides}, in order to study 
later what are the main modifications introduced when taking into
account the dissipative dynamics present during inflation \cite{br4,br}. 
We consider the standard superpotential for the supersymmetric
hybrid inflationary model, Eq. (\ref{superpot}).
Without taking into account SUSY 
breaking, the zero-energy global minimum in the model
Eq. (\ref{superpot}) is located at the vacuum expectation values (VEV)
$S$=0, $\Phi_1=\Phi_2= \mu$. If the gauge group $G$ 
of $\Phi_1$ and $\Phi_2$
is identified with a GUT symmetry,
the scale $\mu$ would be the GUT symmetry breaking scale.  
On the other hand, for $|S| > \mu$, there is 
a local minimum at $\Phi_1=\Phi_2= 0$ with potential
energy given by the constant term $\kappa^2 \mu^4$. Inflation
occurs while the system is located in this false
vacuum. Here, the inflaton scalar field $S$ and its fermionic partner
remain massless at tree level, while the scalars $\Phi_1$ and
$\Phi_2$ combine into a pair of real scalar and pseudoscalar 
particles with mass $m_+^2=
\kappa^2( |S|^2 + \mu^2)$, and another pair with mass $m_-^2=
\kappa^2( |S|^2 - \mu^2)$; their fermionic superpartners are degenerate with
mass $m_F= \kappa |S|$. 

Due to the splitting in the masses, there is a
non-vanishing one loop radiative correction to the potential, $\Delta
V$, which provides the necessary slope and mass correction\footnote{We
will not add any additional SUSY breaking mass term for the inflaton during
inflation, so that its mass is given purely by the radiative
corrections. Given that the $\mu$ scale is typically of the
order of the GUT scale, SUSY breaking masses of the order of $O(1)$
TeV are negligible unless the coupling $\kappa < O(10^{-5})$.} to the
inflaton potential to drive slow-roll inflation. In particular, the first and
second derivatives of the effective inflaton potential are given
respectively by: 
\bea
\Delta V^\prime&=& \frac{\sqrt{2}\kappa^4 \mu^3 {\cal N}}{16
  \pi^2}\left( x^3 \ln 
\frac{(x^2-1)(x^2+1)}{x^4}+ x \ln \frac{x^2+1}{x^2-1} \right)
= \frac{\sqrt{2}\kappa^4 \mu^3 {\cal N}}{16 \pi^2} F_1[x] \,, \label{delvp}\\
\Delta V^{\prime\prime }&=& \frac{\kappa^4 \mu^2 {\cal N}}{16
  \pi^2}\left( 3 x^2 \ln 
\frac{(x^2-1)(x^2+1)}{x^4}+ \ln \frac{x^2+1}{x^2-1} \right)
= \frac{\kappa^4 \mu^2 {\cal N}}{16 \pi^2} F_2[x] \,,\label{delvpp}
\eea
where ${\cal N}$ is the dimensionality\footnote{We take ${\cal N}=1$
throughout this paper unless otherwise explicitly stated.} of the $\Phi_{1,2}$
representations, $x \equiv \phi_S/(\sqrt{2} \mu)$, $\phi_S$ is the real part of the
complex field $S$, and we are setting the imaginary components of all
the fields to zero for simplicity. For large $x$, we have: $F_1[x]
\sim 1/x$, $F_2[x] \sim -1/x^2$. 

Standard  (cold) ``slow-roll'' inflation is characterized by having small slow-roll
parameter $\epsilon_H \ll 1$ and $\eta_H \ll 1$, with 
\bea
\epsilon_H &=&
\frac{m_P^2}{2}\left(\frac{V^\prime}{V}\right)^2=\frac{\kappa^2}{(4
  \pi)^2} \left( \frac{\kappa m_P}{4 \pi  
  \mu}\right)^2 ({\cal N} F_1[x]) ^2 \approx \frac{\kappa^2}{(4 \pi)^2} \left(
\frac{\kappa m_P}{4 \pi \mu}\right)^2 \frac{{\cal N}^2}{x^2} \label{epsh}\,,\\
\eta_H &=& m_P^2 \frac{V^{\prime \prime}}{V}=\left( \frac{\kappa m_P }{4 \pi\mu}\right)^2 {\cal N} F_2[x]\approx
-\left( \frac{\kappa m_P}{4 \pi\mu}\right)^2 \frac{{\cal N}}{x^2} \label{etah}
\,, 
\eea
where $m_P=2.4\times 10^{18}$ GeV is the reduced Planck mass. 
Therefore, during inflation the evolution of the inflaton field
is well approximated by
\be
3 H \dot \phi_S + V^\prime \simeq 0 \,.
\label{eom}
\ee
At this point one should worry about supergravity (SUGRA) corrections
to the  inflaton potential. Generically, those give rise to mass
corrections for the scalars during inflation of the order of $O(H^2)$,
with $\eta_H \sim O(1)$, spoiling inflation (the so-called $\eta$
problem \cite{copeland, etaproblem}).  However, such corrections are
not present if we take the minimal Kahler potential for the fields with a
superpotential like Eqs.(\ref{superpot}) and (\ref{superpot2}), and this is the
choice we adopt in this paper. Nevertheless, they will induce a
quartic term (plus some higher order corrections) in the inflaton potential
\cite{copeland,linderiotto,kawasaki,shafi}, with  
\be
V \simeq \kappa^2 \mu^4 ( 1 + \frac{\phi_S^4}{8 m_P^4} + \cdots)+ \Delta V  \,.
\label{Vsugra} 
\ee 
The quartic term  dominates the inflationary dynamics when
$\phi_S \sim m_P$, which happens for $\kappa \sim O(1)$.

The values of the coupling $\kappa$ and the scale $\mu$ consistent
with the inflationary dynamics are obtained by demanding that (a) we
have ``enough'' inflation (at least 60 e-folds), and (b) that the
amplitude of the primordial spectrum generated by the inflaton vacuum 
fluctuations  are in the range given by COBE observations. The former
constraint gives the value of the inflaton field $N_e \,(\simeq 60)$
e-folds before the end of inflation,
\be
x_{N} \approx \sqrt{2 N_e} \left(\frac{\kappa m_P}{4 \pi\mu} \right)\,,
\label{xN}
\ee
which is then used to evaluate the amplitude of the primordial curvature
spectrum,
\be
P^{1/2}_{\cal R} = \left(\frac{H}{\dot
  \phi_S}\right) \left(\frac{H}{2 
\pi}\right)= \sqrt{\frac{2}{3}}\left(\frac{\kappa}{4\pi}\right)^2
\left(\frac{4 \pi \mu}{\kappa m_P}\right)^3 \frac{1}{F_1[x_{N}]}\approx
\sqrt{\frac{4 N_e}{3}} \left(\frac{\mu}{m_P}\right)^2 \,.
\label{spectrum0}
\ee 
Therefore, using the COBE normalization 
\cite{COBE,WMAP} $P^{1/2}_{\cal R} = 5 \times 
10^{-5}$ at $N_e\simeq 60$, we have\footnote{Corrections to
this estimation appear for small values of $\kappa < 0.01$, for which
$x_N \sim O(1)$.}  that $\mu \simeq 2\times 10^{-3} m_P \simeq 
5\times 10^{15}$ GeV. 

Given that implicitly we are working with a SUGRA model, at most the
VEV of the inflaton field could be of the order of the Planck scale,
preferably below that scale. From Eq. (\ref{xN}) one can see that
$\phi > m_P$ for values of $\kappa > 0.8$ and so are excluded
\cite{linderiotto}. Moreover, taking into account the quartic
SUGRA correction in Eq.(\ref{Vsugra}), 
the spectrum becomes blue tilted ($n_S > 1$)
\cite{panag,linderiotto,kawasaki,shafi} already for 
$\kappa \simeq 0.05$, 
which is not favored by the observational data on the spectral index
from WMAP\footnote{From the WMAP data only, the best fit value 
  without running of the spectral index is $n_S= 0.99 \pm 0.04$. A
  recent analysis of the combined Lyman-$\alpha$ forest spectra
+ WMAP \cite{lyman} also gives $n_S= 0.99 \pm 0.03$  with no running,
and $n_S= 0.959 \pm 0.036$ with $d n_S/ d \ln k = -0.033 \pm
0.025$. However, a similar analysis in Ref. \cite{seljak} gives $n_S= 0.98 \pm 0.02$ with $d n_S/ d \ln k = -0.003 \pm 0.01$.}, 
$n_S= 0.93 \pm 0.03$ \cite{WMAP}.

\section{Dissipative dynamics during inflation}
\label{sect3}

Dissipative effects can be important  already during inflation, modifying
the inflationary dynamics described by Eq. (\ref{eom}). These are
related to the quantum corrections in  the effective
potential of the background field. When neither the decay of the
inflaton  nor that of the fields coupled to the inflaton are
kinematically allowed, loop corrections to the propagators are real,
and they are  absorbed into the renormalized masses and couplings,
order by order in perturbation theory. On the other hand, when the
inflaton (or the fields coupled to the inflaton) can decay into
other particles, the propagators in the loop have the standard Breit-Wigner
form, with an imaginary contribution related to the decay rate
$\Gamma$. 
Therefore, when computing the 1-loop effective potential for the inflaton
field, the contributions associated to the decay rate
lead to dissipative effects \cite{br4,br}.  In general this reflects itself
in the form of temporally nonlocal terms in the inflaton evolution
equation.  Under certain
approximations this translate into a simple
effective friction term $\Upsilon_S$ in the
equation of motion for the background inflaton field \cite{br4,br},
\be
\ddot \phi_S + (3 H + \Upsilon_S) \dot \phi_S + \Delta V^\prime =0 \,.
\label{ddotphi}
\ee
The emergence of this friction term due to these
underlying decay channels implies the dynamics of the
system is such that part of the inflaton energy is dissipated into the lighter
particles produced in the decays, i.e., into radiation $\rho_R$, with
\be
\dot \rho_R + 4 H \rho_R = \Upsilon_S \dot \phi_S^2 \,.
\label{dotrhor}
\ee
Although the basic idea of interactions leading to dissipative effects
during inflation is generally valid, the above set of equations
has strictly been derived in \cite{br4,br}
only in the adiabatic-Markovian
limit, i.e., when the fields involved are moving slowly, which
requires 
\be
\frac{\dot \phi_S}{\phi_S} < H < \Gamma \,,
\ee
with $\Gamma$ being the decay rate. The second inequality, $ H < \Gamma$
is also the condition for the radiation (decay products) to thermalize.   

Thus in general any inflation model could have two very distinct 
types of inflationary dynamics, which have been termed cold and
warm \cite{wi,br}.  The cold inflationary regime
is synonymous with the standard inflation picture
\cite{oldi,ni,ci}, in which dissipative effects are completely
ignored during the inflation period.  On the other hand,
in the warm inflationary
regime dissipative effects play a significant role in 
the dynamics of the system.  A rough quantitative measure
that divides these two regimes is
$\rho_R^{1/4} \approx H$, where
$\rho_R^{1/4} > H$ is the warm inflation regime 
and $\rho_R^{1/4} \stackrel{<}{\sim} H$
is the cold inflation regime. This criteria is independent of
thermalization, but if such were to occur, one sees this criteria
basically amounts to the warm inflation regime corresponding to when $T
> H$. This is easy to understand since the typical inflaton mass during
inflation is $m_\phi \approx H$ and so when $T>H$, thermal fluctuations
of the inflaton field
will become important. This criteria for entering the warm inflation
regime turns out to require the dissipation of a very tiny fraction of
the inflaton vacuum energy during inflation. {}For example, for
inflation with vacuum ({\it i.e.} potential) energy at the GUT scale
$\sim 10^{15-16} {\rm GeV}$, in order to produce radiation at the scale
of the Hubble parameter, which is $\approx 10^{10-11} {\rm GeV}$, it just
requires dissipating one part in $10^{20}$ of this vacuum energy density
into radiation. Thus energetically not a very significant amount of
radiation production is required to move into the warm inflation regime.
In fact the levels are so small, and their eventual effects
on density perturbations and inflaton evolution are so significant, that
care must be taken to account for these effects in the analysis of any
inflation models.

The conditions for slow-roll inflation ($ \dot \phi_S^2 \ll V$, $\ddot
\phi_S \ll H \dot \phi_S$) are modified in the
presence of the extra friction term $\Upsilon_S$, and we have now:
\bea
\epsilon_\Upsilon &=& \frac{\epsilon_H}{(1+r)^2} \label{epsups}\,,\\
\eta_\Upsilon &=& \frac{\eta_H}{(1+r)^2} \label{etaups} \,,
\eea
where $r= \Upsilon_S/(3 H)$, and $\epsilon_H$, $\eta_H$ are 
the slow-roll parameters without dissipation given in
Eqs. (\ref{epsh}) and (\ref{etah}).  In addition, when the friction term
$\Upsilon_S$ depends on the value of the inflaton field, we can define
a third slow-roll parameter
\be
\epsilon_{H \Upsilon} = \frac{r}{(1+r)^3} \beta_\Upsilon  \label{epsupsi}\,,
\ee
with
\be
\beta_\Upsilon = \frac{V^\prime}{3 H^2}\frac{
\Upsilon_S^\prime}{\Upsilon_S}\,. 
\ee
Similarly to the slow-roll regime without dissipation, when
$\eta_\Upsilon < 1$, $\epsilon_\Upsilon < 1$, 
and $\epsilon_{H\Upsilon} < 1$, Eqs. (\ref{ddotphi}) and (\ref{dotrhor})
are well approximated by:
\bea
\dot \phi_S &\simeq& -\frac{\Delta V^\prime}{3 H}\frac{1}{1+r}\,, \\
\rho_R &\simeq& \frac{\Upsilon_S}{4H} \dot \phi_S^2 \simeq
\frac{1}{2}\frac{r}{(1+r)^2} \epsilon_H V\,,
\eea
and the number of e-folds is given by:
\be
N_e \simeq - \int_{\phi_{Si}}^{\phi_{Se}} \frac{3 H^2}{ \Delta V^\prime} (
1 + r) d \phi \,. \label{Neups} 
\ee
Obviously, when $\Upsilon_S \ll 1$ we recover the standard ``cold'' 
hybrid inflation (CHI) scenario. 

The effect of the dissipative term is twofold: on one hand,    
dissipation of the vacuum energy into radiation acts as an extra
friction term and  slows down the
motion of the inflaton field, so that inflation last longer. That means
that when $\Upsilon_S$ is non negligible, we would require in general
smaller initial values of the inflaton field 
in order to have ``enough'' (at
least 60 e-folds) inflation, Eq. (\ref{Neups}). 
On the other hand,  fluctuations in the
radiation background affect those of the inflaton field through the
interactions, and this in turn will affect the primordial spectrum
generated during inflation. Approximately, one can say that when 
$T > H$ the fluctuations of the inflaton field are induced by the
thermal fluctuations, instead of being vacuum fluctuations, with a
spectrum proportional to the temperature of the thermal bath. 
We notice that having $T>H$ does not necessarily require $\Upsilon_S >
3H$. Dissipation may  not be strong enough to alter the dynamics of the
background inflaton field, but it can be enough even in the weak
regime to affect its fluctuations, and therefore the
spectrum. Depending on the different regimes,
the spectrum of the
inflaton fluctuations $P^{1/2}_{\delta \phi}$ is given for
cold inflation \cite{Guth:ec}, weak
dissipative warm inflation \cite{Moss:wn,Berera:1995wh} 
and strong dissipative warm inflation \cite{Berera:1999ws}
respectively by
\bea
T < H:  && P^{1/2}_{\delta \phi}|_{T=0}\simeq \frac{H}{2 \pi} \,, \\
\Upsilon_S < H < T: && P^{1/2}_{\delta \phi}|_{T}
 \simeq \sqrt{TH} \sim
\sqrt{\frac{T}{H}}P^{1/2}_{\delta \phi}|_{T=0} \,, \\ 
\Upsilon_S > H: && P^{1/2}_{\delta \phi}|_{\Upsilon} \simeq
\left(\frac{\pi \Upsilon_S}{ 4 H}\right)^{1/4}\sqrt{T H}\sim 
\left(\frac{\pi \Upsilon_S}{ 4 H}\right)^{1/4}\sqrt{\frac{T}{H}}
P^{1/2}_{\delta \phi}|_{T=0} \,,   
\,,   
\eea
with the amplitude of the primordial spectrum of the curvature
perturbation given by:
\be
P^{1/2}_{\cal R} = \left|\frac{H}{\dot
  \phi_S}\right| P^{1/2}_{\delta \phi} \simeq
\left|\frac{3 H^2}{\Delta V^\prime}\right| (1+r) P^{1/2}_{\delta \phi}
\label{spectrumr} \,.
\ee
Given the different ``thermal'' origin of spectrum, the spectral index
also changes with respect to the cold inflationary scenario
\cite{arjunspectrum,hmb1,warmspectrum,warmrunning}, even in the weak
dissipative warm inflation regime when the evolution of the inflaton field is
practically unchanged. Again, for the different regimes, it is obtained: 
\bea
T < H: && n_S -1 = -6 \epsilon_H + 2 \eta_H \,,\\
\Upsilon_S < H < T: &&  n_S-1= -\frac{17}{4} \epsilon_H + \frac{3}{2}
\eta_H - \frac{1}{4} \beta_\Upsilon \,, \\
\Upsilon_S > H : && n_S-1= (-\frac{9}{4} \epsilon_H + \frac{3}{2}
\eta_H - \frac{9}{4} \beta_\Upsilon)\frac{1}{(1+r)}  \,.
\eea
In the latter case there could be appreciable departures from scale
invariance in the spectrum; we notice again than in the strong
dissipative case, slow-roll only demands $\epsilon_H \ll (1 +r)^2$ and
$\eta_H \ll (1+r)^2$, whereas the spectral index depends on the ratios
$\epsilon_H/(1+r)$ and $\eta_H/(1+r)$, which not necessarily are much
smaller than 1.   

The question then is not whether there is dissipation during inflation, but
whether this will affect the inflationary predictions. First, how large
can $\Upsilon_S$ be in a realistic set-up. 
In the calculations in \cite{br4,br,hm} a robust mechanism 
for dissipation during inflation has been identified.
The basic interaction structure for this mechanism is
\begin{equation}
{\cal L}_I = - \frac{1}{2}g^2 \phi^2 \chi^2 -
g' \phi {\bar \psi_{\chi}} \psi_{\chi} - h \chi {\bar \psi_d}\psi_d ,
\label{lint}
\end{equation}
where $\phi$ is the inflaton field, $\chi$ and
$\psi_{\chi}$ are additional fields to which the inflaton couples, and
$\psi_d$ are light fermions into which the scalar $\chi$-particles can
decay $m_\chi > 2m_{\psi_d}$.  This interaction structure can
be identified in both models we are studying in
this paper Eqs. (\ref{superpot}) and (\ref{superpot2}).
In the next two subsections, the dissipative properties of
these two models will be computed based on the results
on \cite{br4,br} and then the effect of this dissipation
to inflation will be studied.

\subsection{Decay into massive fermions:} 
\label{sect3a}

First consider the model Eq. (\ref{superpot}), which
has only the minimal matter content.
In this model, the scalar with the largest
mass $m_+$ can decay into its fermionic superpartner $\tilde{\Phi}_+$
and a massless inflatino, with decay rate\footnote{The interaction
Lagrangian is given by ${\cal L}= 
- (\kappa/\sqrt{2}) \phi_+ \bar{\tilde{S}}\tilde{\Phi}_+$.}:
\begin{equation}
\Gamma_+= \frac{\kappa^2}{16 \pi} m_+ \left(\frac{1}{x_{N}^2 +
  1}\right)^2 \,, 
\end{equation}
where again $x_{N}$ is the value of the inflaton field (normalized by
$\mu$) $N_e$ efolds before the end of inflation. This decay rate is always 
smaller than the rate of expansion during inflation:
\begin{equation}
\frac{\Gamma_+}{H}= \sqrt{3}\frac{\kappa^2}{16 \pi}
\frac{m_P/\mu}{(x_N^2+1)^{3/2}} \ll 1 \,,
\end{equation}
and strictly speaking the adiabatic-Markovian approximation would not
apply. Nevertheless, in order to get some numbers, let us proceed and estimate
the dissipative coefficient and the amount of radiation produced. The
former is given by:
\begin{eqnarray}
\Upsilon_S(\phi_S) &=& \frac{\sqrt{2} (\kappa^4/4) (\Gamma_{+}/m_+)}
{64\pi \sqrt{1 + (\Gamma_{+}/m_{+})^2}
\sqrt{\sqrt{1 + (\Gamma_{+}/m_{+})^2}+1}} \frac{\phi_S^2}{m_+}
\nonumber \\
&\simeq&\frac{\pi^3}{2}\left(\frac{\kappa}{4 \pi}\right)^5 
\frac{ x_N^2 }{( x_N^2 +1)^{5/2}}\mu 
\simeq 
\frac{\pi^3}{2}\left(\frac{\kappa}{4 \pi}\right)^5 \frac{\mu}{x_N^3} \,,
\end{eqnarray}
which is always suppressed with respect to the expansion rate during
inflation, $H \simeq \kappa \mu^2/(\sqrt{3}m_P)$:
\begin{equation}
\frac{\Upsilon_S}{3 H} \simeq \frac{\pi^2}{8 \sqrt{3}}\left(\frac{\kappa}{4
  \pi}\right)^4 \frac{m_P}{\mu}\frac{x_N^2}{(x_N^2+1)^{5/2}}\ll 1 \,.
\end{equation}
Nevertheless, the amount of ``radiation'' produced, i.e., the energy density
dissipated from the inflaton, could be larger than $H^4$, 
\bea
\frac{\rho_R}{H^4} &\simeq& \frac{9}{2}\frac{\Upsilon_S}{3 H} \epsilon_H 
\frac{m_P^4}{\kappa^2 \mu^4}
\simeq \frac{9}{256 \sqrt{3}}\left(\frac{\kappa}{4 \pi}\right)^6
\left(\frac{m_P}{\mu}\right)^7 \frac{x_N^2}{(x_N^2+1)^{5/2}} F_1[x_N]^2 \,.
\eea
Given that the ratio $\rho_R/H^4$ goes like the inverse of $x_N^5$, the
ratio increases as the value 
of the inflaton field decreases during inflation. However, it only
becomes larger than one for $\kappa \sim 0.1$ toward the end of 
inflation, well after the 60 e-folds before the end. 
We can conclude then that  dissipation in this example is too weak to 
affect either the spectrum of the primordial perturbations or the
dynamics of the inflaton field.


\subsection{Decay into massless fermions:}
\label{sect3b}

As we have seen, dissipation through the decay into massive fermions
does not have  much effect during inflation. However, in more
realistic models, one  would expect the presence of other fields,
which in principle are not directly relevant during inflation but can
play a r\^ole during/after reheating. 
For example, fields coupled to
either $\Phi_1$ or $\Phi_2$  are massless during inflation, and
become massive at the global  minimum. 
Thus the model Eq. (\ref{superpot2}), where we have introduced a pair of
matter fields, $\Delta$ and $\bar  \Delta$, coupled to $\Phi_2$. 
Because they are massless during inflation, they do not contribute to
the 1-loop effective potential, and radiative corrections are the same
as in the previous case, with the slope and 
curvature of the effective potential given by Eqs. (\ref{delvp}) and
(\ref{delvpp}).   
But now the  heaviest field with mass $m_+$ can decay into the
massless fermionic partners of $\Delta$ and $\bar  \Delta$, with decay rate:
\begin{equation}
\Gamma_+= \frac{g^2}{16 \pi} m_+ \,. 
\end{equation}
Since now there is no phase space suppression factor in the decay
rate, we have  $\Gamma_+ \propto \phi_S$, and this can be much larger
than the Hubble rate during inflation: 
\begin{equation} 
\frac{\Gamma_+}{H}= \sqrt{3}\frac{g^2}{16 \pi}
\left(\frac{m_P}{\mu}\right) (x_N^2+1)^{1/2}\,. 
\end{equation}
Having $\Gamma_+/H>1$, all the way up to the end of inflation, only
requires $g > 0.16$ for $\kappa <0.001$ ( $g > 0.01$ for
$\kappa =0.5$). This allows us to apply the
adiabatic-Markovian limit in the effective equation of motion for the inflaton
background field, Eq. (\ref{ddotphi}),
with the dissipative coefficient given now by:
\begin{equation}
\Upsilon_S \simeq \frac{\pi^2}{2} \left(\frac{\kappa}{4
  \pi}\right)^3 \left(\frac{g^2}{16 \pi}\right) \frac{x_N^2}{(1 +
  x_N^2)^{1/2}} \mu \label{upsilon} \,,
\end{equation} 
and the ratio to the Hubble rate is given by: 
\begin{equation}
\frac{\Upsilon_S}{3 H} \simeq \frac{\kappa^2}{128 \sqrt{3}\pi} 
\left(\frac{g^2}{16\pi}\right)\frac{x_N^2}{(1 +
  x_N^2)^{1/2}} \frac{m_P}{\mu} \,, 
\end{equation}  
which behaves like  $\Upsilon_S/(3H) \propto x_N\propto \phi_S$, and
so decreases  during inflation. 
That is, the evolution of
the inflaton field may change 
from being dominated by the friction term $\Upsilon_S$ to be dominated
by the Hubble rate $H$. Whether the transition between these two
regimes happens before or after 60 e-folds will depend on the value of
the parameters of the model like $\kappa$ and $g$. The amount of
``radiation'' obtained through the dissipative term,  
is given by:
\bea
\frac{\rho_R}{H^4} &\simeq& \frac{9}{2}\frac{r}{(1+r)^2} \epsilon_H 
\frac{m_P^4}{\kappa^2 \mu^4} \label{rhorh4}\,, 
\eea
which even when $\Upsilon_S < H$ could give rise to a thermal bath
with $T > H$. In particular, we can have: 

(a) $\Upsilon_S > 3 H$, and $T > H$  ($\dot \phi_S \simeq
-V_\phi/ \Upsilon_S$):  
\be
 \frac{\rho_R}{H^4} \simeq
 \frac{36 \sqrt{3}}{\pi^2}\frac{1}{g^2}
 \left(\frac{m_P}{\mu}\right)^5 \frac{1}{x_N^3}\,,
\ee

(b) $\Upsilon_S < H$ ($\dot \phi_S \simeq -V_\phi/ (3 H)$):
\be
 \frac{\rho_R}{H^4} \simeq
\frac{9}{256 \sqrt{3} \pi}\left(\frac{\kappa}{4 \pi}\right)^4 
 \left(\frac{g^2}{16 \pi}\right)
\left(\frac{m_P}{\mu}\right)^7 \frac{1}{x_N}
 \,.
\ee

\begin{figure}[t] 
\hfil\scalebox{0.5} {\includegraphics{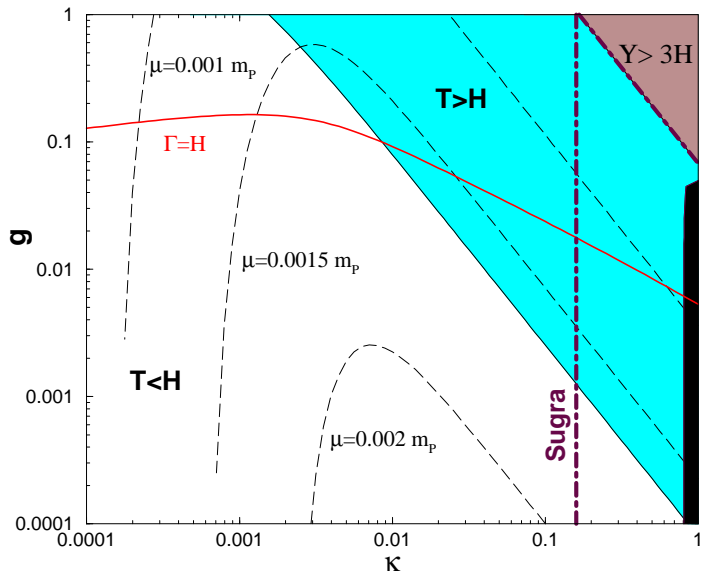}}\hfil
\caption{\label{plot1} 
Regions of cold ($T< H$), and warm ($T > H$) SUSY hybrid
inflation in the $\kappa-g$ plane. The warm inflation region is
divided into the weak dissipative regime with $T > H$ and $\Upsilon_S <
3H$ (lighter shaded region), and the strong dissipative regime with
$\Upsilon_S > 3H$ (darker shaded region). Included are also the
contour plots of constant $\mu$, and the adiabatic-Markovian limit
$\Gamma_+= H$. The black region on the right of the plot is excluded
because $\phi_S > m_P$. In addition, when SUGRA corrections are
taking into account, values to  the left and down 
the wide dot-dashed line are excluded.}
\end{figure}

The values of the couplings $\kappa$ and $g$ for which we
could have cold or warm inflation, and strong or weak dissipative
dynamics, are plotted in Fig. (\ref{plot1}). In order to get the
different regions, we have proceeded as follow: for each pair of
values in the plane $\kappa-g$, the value of the inflaton field at the
end of inflation is determined.  This is done in the cold and weak dissipative 
regimes by the condition\footnote{The value of $\eta_\Upsilon$ becomes
larger than 1 before the other two slow-roll parameters.}
$\eta_\Upsilon=1$, Eq.(\ref{etaups}). In the strong dissipative regime
inflation can end either with $\eta_\Upsilon =1$ or
it may also happen that most of the vacuum energy is already transferred into
radiation during inflation, and then inflation will end when
$\rho_R \simeq \kappa^2 \mu^4$ instead. 
In this case, whichever occurs first fixes the value of
the inflaton field at the end of inflation.   
The value of the inflaton field
at 60 e-folds of inflation is then obtained from
Eq. (\ref{Neups}). This in turn fixes the value of the dissipative
coefficient $\Upsilon_S$, Eq. (\ref{upsilon}), the temperature of thermal
bath, Eq. (\ref{rhorh4}), and therefore the amplitude of the spectrum
Eq. (\ref{spectrumr}).  The COBE normalization is then used to fix the
value of the scale $\mu$.     
In order to match the expressions for the spectrum across 
the different regimes, we have used a simple expression with :
\be 
P^{1/2}_{\cal R} = 
\left|\frac{3 H^2}{\Delta V^\prime}\right| (1+r) \left( 1 +
 \sqrt{\frac{T}{H}} \right) \left( 1 +\left(\frac{\pi
   \Upsilon_S}{4H}\right)^{1/4} \right)\left(\frac{H}{2 \pi}\right) \,.
\ee 

We can see in Fig. (\ref{plot1}) that the strong dissipative regime
$\Upsilon_S > 3H$ requires large values of the couplings, $\kappa \sim g
\sim O(1)$; for values $\kappa \simeq g \simeq 0.1$ we are in the weak
dissipative regime; and for values $\kappa \simeq g \simeq 0.01$ we
recover the cold inflationary scenario.  
Typically, for a fixed value of the scale $\mu$ the amplitude of the
spectrum in the strong dissipative regime would be larger than the one
generated at zero $T$. The COBE normalization implies then a smaller
value of the inflationary scale 
$\mu$. For example, for $\kappa = g =1$ we have $\mu\simeq 10^{13}$ GeV,
whereas pushing the coupling toward its perturbative limit, $\kappa
=g= \sqrt{4 \pi}$ we would get $\mu \simeq 2 \times 10^{10}$ GeV. 
On the other hand, going from the cold to the weak dissipative regime,
the value of $\mu$ only varies by a factor of 2 or 3, and it is still
in the range of the GUT scale $O(10^{15})$ GeV. 

As mentioned before, the quartic term in the inflaton potential
induced by SUGRA corrections becomes non negligible for not very small
values of $\kappa$. In the CHI scenario,
the value of the inflaton 
field becomes larger than $m_P$ for $\kappa \geq 0.15$, and
consequently that region is excluded. The same constraint applies also
in the weak dissipative scenario. However, in the strong dissipative
regime, with $\Upsilon_S > 3H$, the extra friction term keeps the
value of the inflaton field below the Planck scale, and the constraint
on $\kappa$ can be avoided. 

\begin{figure} 
\hfil\scalebox{0.5}{\includegraphics{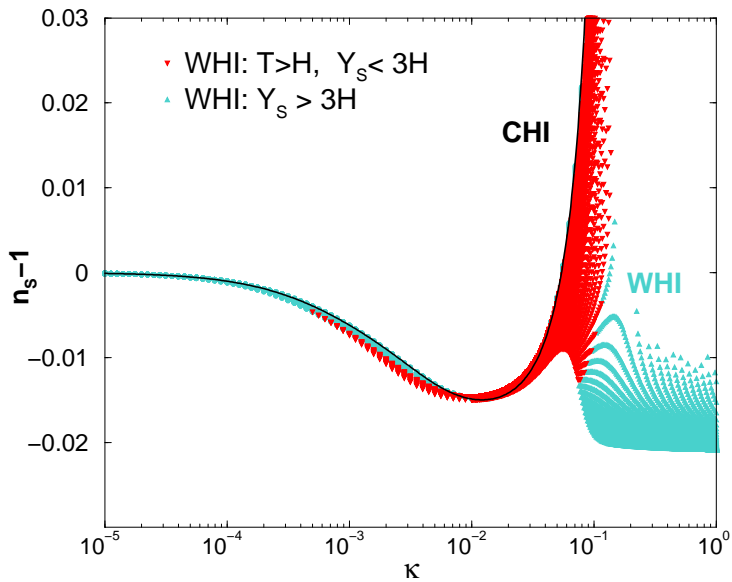}}\hfil
\caption{\label{plot2}
Spectral index for cold SUSY hybrid inflation (solid line, CHI),
and warm inflation (gray region, WHI). The weak dissipative regime ($T >
H$ but $\Upsilon_S < 3H$) is given by the darker gray region
(triangle down);the  strong dissipative regime ($\Upsilon_S > 3H$) 
is given by the light gray region 
(triangle up). }
\end{figure}

In Fig. (\ref{plot2}) we have compared the prediction for the spectral
index of the scalar spectrum of perturbations in both the CHI
scenario, and warm hybrid inflation (WHI). From the warm inflation
scenario we can always recover the CHI prediction by taking $g \ll 1$. In
standard SUSY GUT hybrid inflation, for small values of the coupling
$\kappa$ the spectrum is practically scale invariant, it reaches a
minimum around $\kappa \simeq 0.01$, and then rises due to SUGRA
corrections up to positive values, which are disfavoured by WMAP
results. But in the weak and the strong
dissipative regime, due to the different origin of the spectrum, we
get that the spectral index is still below 1 even for values of the coupling 
$\kappa > 0.01$. This is especially true in the strong dissipative
regime, where the 
dynamic is such that the inflaton field is well below the Planck
scale and SUGRA corrections are negligible. In that regime the
departure from scale invariance is within the observational value,
with $n_S -1 \simeq -0.022$.

\vspace{0.2cm}

\begin{figure} 
\hfil\scalebox{0.5}{\includegraphics{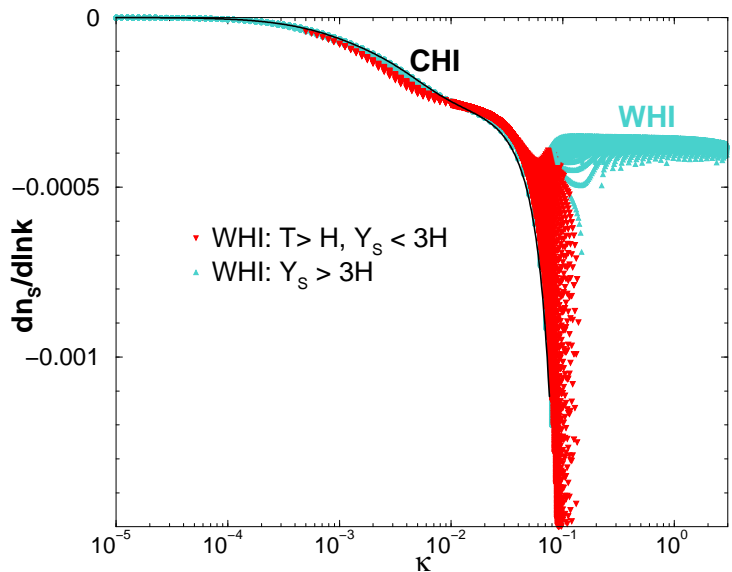}}\hfil
\caption{ \label{plot3}
Running of the spectral index for cold SUSY hybrid inflation
(solid line, CHI), and warm hybrid inflation (gray region, WHI).
The weak dissipative regime ($T >
H$ but $\Upsilon_S < 3H$) is given by the darker gray region
(triangle down); the strong dissipative regime ($\Upsilon_S > 3H$) 
is given by the light gray region 
(triangle up).} 
\end{figure}

In Fig. (\ref{plot3}) we have also compared the running of the
spectral index, $dn_S/d \ln k$ in both scenarios. The running although
negative is much smaller than the value preferred by 
the WMAP data, $d n_S/d \ln k \simeq -0.031^{+0.016}_{-0.015}$. 
 Again, in the strong dissipative regime we can have larger values of
 the couplings $\kappa$ and $g$ compatible with observations, but the
 predicted running of the spectral index is still small, with 
$d n_S/d \ln k \simeq -4\times 10^{-4}$. Nevertheless, it is not
yet clear the statistical relevance of this result, which in any case
would be finally confirmed or excluded by the Planck Satellite experiment.  

\vspace{0.2cm}

\begin{figure} 
\hfil\scalebox{0.5}{\includegraphics{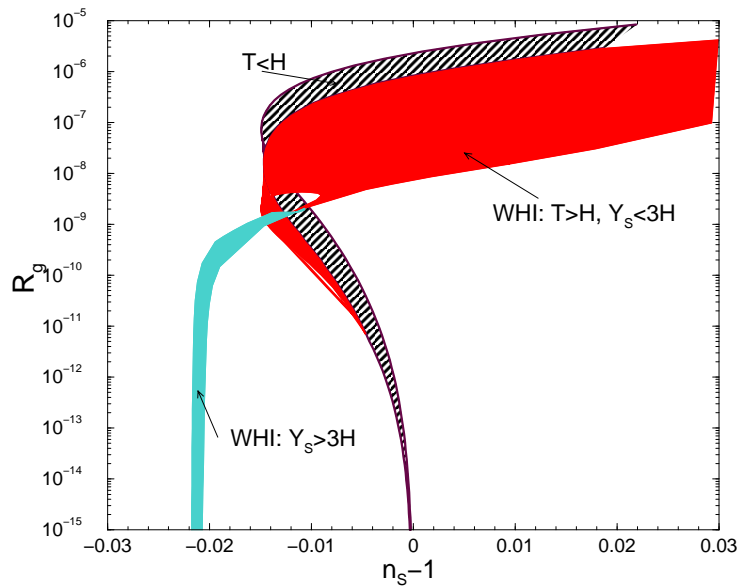}}\hfil
\caption{ \label{plot4}
Tensor-to-scalar ratio versus the predicted scalar spectral index
$n_S-1$. The cold regime, $T<H$, is given by 
the dark grey areas; the weak
dissipative regime, $T>H$ but $\Upsilon_S < 3H$, is given by the
striped area; and the strong dissipative regime, $\Upsilon_S > 3H$, 
is given by the light gray region.} 
\end{figure}

Concerning the primordial spectrum of tensor perturbations, as they
do not couple strongly to the thermal background, the amplitude is the
same than in CHI, with
\be
P_{tensor}= \frac{2}{m_P^2}\left(\frac{H}{2 \pi}\right)^2\,.
\label{ptensor}
\ee
As we have seen, strong dissipation  ($\kappa\sim O(1)$) translates
into a smaller inflationary scale $\mu$ compared to that of standard
CHI, therefore a lower value of $H$ and a smaller contribution of the
gravitational waves to the spectrum. However, the same level of
primordial tensor perturbations can be obtained decreasing the value of
$\kappa$ and no dissipation. This can be seen in Fig. (\ref{plot4}),
where  we have plotted the prediction for the tensor-to-scalar ratio, 
defined as  
\be
R_g=\frac{P_{tensor}}{P_{\cal R}} \,,
\ee
versus the prediction for the scalar spectral index $n_S-1$, for the
different regimes (cold, weak dissipation, and strong 
dissipation). The smallest
values of $\kappa$ correspond to a practically scale invariant
spectrum. At present, from the Cosmic Microwave Background polarization measurements the
tensor-to-scalar ratio is poorly constraint ($R_g < 0.4$), although in
the future 
ratios as low as $10^{-6}$ could be probed \cite{tensor}. 
Still, such
a gravitational background can be achieved (but not larger) in this
kind of models only 
in the cold or weak dissipative regime, and typically  for values of
the coupling $\kappa$ close to the maximum allowed by SUGRA
corrections (blue tilted spectrum). 
On the other hand, in the strong dissipative regime we
obtain a clear prediction that distinguishes this regimes from the
others: no expected tensor signal, with $R_g < 10^{-9}$, and a red
tilted spectrum with $n_S\simeq 0.98$. 

\section{Reheating Temperature} 
\label{sect4}

One of the main constraints on model building of inflationary
supersymmetric particle physics models comes from the gravitino
constraint on the reheating $T$ after inflation. Gravitinos
with a typical mass of the order of $O(1-100)$ TeV can be thermally
produced during the radiation dominated era that follows inflation. If
$T$ is too high, we will have too many of them, and 
their subsequent decay will interfere with the predictions of the
abundances of light elements at the time of big bang nucleosyntheses
(BBN) \cite{gravitino}. This puts an upper bound on the reheating $T_{RH}$
typically of the order of $10^{9}$ GeV \cite{gravitino}, and a more
recent analysis on BBN has lowered this bound to  $T_{RH} \lessa 10^{7}
GeV$ for a gravitino  mass $m_{3/2} \simeq O(1)$ TeV
\cite{gravitino2,gravitino3}.  
In cold inflation, the reheating $T$ 
after inflation can be well approximated by:
\be
T_{RH} \simeq \left(\frac{90}{\pi^2 g_*}\right)^{1/4}
\sqrt{\Gamma_\phi m_P} \,, 
\ee
where $\Gamma_\phi$ is the decay rate of the inflaton field, and $g_*$
the effective number of degrees of freedom (typically of the order of
O(100) in a SUSY model). Therefore, the reheating constraint translates
in the inflaton not decaying too fast after inflation, which may
implies some further constraint on the couplings\footnote{This can be
  avoided if the gravitino abundance is diluted by the entropy
  produced during the late-decay of some other particle \cite{kohri}.}. 

In this letter we have just minimally
extended the inflationary sector  by adding a pair of extra matter
fields $\Delta$, $\bar \Delta$, with generic coupling $g$. This allows
the decay of the $\Phi_i$ fields 
already during inflation, and the possibility of having a $warm$ regime
of hybrid inflation, when the couplings are not very small, say
$\kappa, g > 0.1$. The question then is whether this large coupling $g$ 
could give rise to a large decay rate, and therefore a too large
$T_{RH}$. From the $S$, $\Phi_i$ sector we would have scalars and 
pseudoscalars with masses $m_S= \sqrt{2} \kappa \mu$ (plus a massless
state if the minimum is along the $D$-flat direction
$\Phi_1=\Phi_2$). On the other hand, the $\Delta$, $\bar \Delta$
sector gives both scalars and fermions with a mass $m_\Delta= \sqrt{2}
g \mu$. In order to avoid the decay of the inflaton into these
fermions, it is enough to require $\kappa < 2 g$, which is just a mild
constraint on the values of the couplings. 

In order to complete the transfer of the energy density into radiation
after inflation, the model has to include the decay of the inflaton
field into other lighter fields, with coupling $h_S$ and $\Gamma_S =
m_S h_S^2 /(8 \pi)$. In the strong dissipative regime ($\kappa> 0.1$)
we have $\mu \simeq O(10^{12}- 10^{14})$ GeV, and demanding $T <
10^{9}$ GeV gives $h_S \leq 5\times 10^{-6}$. A similar constraint is
obtained in the CHI scenario  and the weak dissipative regime, where
now $\mu \sim O(10^{15})$ GeV, but $m_S$ decreases with $\kappa <
0.1$.  

As a well-motivated example, which combines inflation with
leptogenesis and light neutrino masses \cite{lazalepto,shafilepto},
the inflaton can decay into right handed (s)neutrino fields $\nu_{Ri}$
($i$= family index). The decay proceed through the
non-renormalizable coupling $\Phi_1 \Phi_1 \nu_{Ri} \nu_{Ri}$,
with decay rate,
\be
\Gamma_S = \frac{1}{8 \pi} \left(\frac{M_i}{\mu}\right)^2 m_S \,,
\ee
where $M_i$ is the RH (s)neutrino mass.  In the CHI scenario, with $\mu
\simeq O(10^{15})$ GeV, $\kappa \simeq 10^{-2}$, and $m_S
\simeq 10^{13}$ GeV, the gravitino constraint $T_{RH} \leq 10^{9}$ GeV
translates roughly into $M_i\simeq 10^{-3} \mu \sim O(10^{12})
GeV$. Those values are also 
consistent with baryogenesis and light neutrino masses
\cite{shafilepto}. This kind of scenario is also viable in the
warm inflationary regime. Being consistent with the observed baryon
asymmetry and the atmospheric neutrino oscillations
does not directly constraint the value of $\kappa$ but the value of
$m_S \sim 10^{13}$ GeV. 
In the warm inflationary regime the
value of the scale $\mu$ required for successful inflation reduces as
we moved into the strong dissipative regime, $m_S$ is  of the
order of $10^{13}$ GeV for $\kappa \simeq O(1)$, and the gravitino
constraint gives now  
$M_i\simeq 10^{-3} \mu \sim O(10^{10})$ GeV. 
Therefore,
a model of $warm$ inflation and leptogenesis without the need of small
couplings would be viable and compatible with observations, in the
strong dissipative regime. 
 
Nevertheless, in the presence of a thermal bath already during
inflation, one could worry about the value of $T$ at the end of
inflation, specially in the strong dissipative regime. It
would be premature to impose the gravitino constraint directly on that
temperature. Taking into account the decay of the inflaton field, the
entropy production during the reheating phase can dilute the abundance
of the gravitinos thermally produced at the end of inflaton
\cite{turner}. Roughly speaking, the 
entropy dilution factor would be $\gamma =S_{RH}/S_{end} \sim T_{end}/T_{RH}$,
where the subindex ``$RH$'' and ``$end$'' refers to the end of
reheating and the end of inflation respectively. One should study in more
detail the reheating phase after ``warm'' hybrid inflation before
drawing any conclusion, taking into account in addition that
production of gravitinos during reheating does not take place in a
pure radiation dominated universe. Inflation will end before the vacuum
energy has been completely dissipated into radiation, and the singlets
may still oscillate around the global minimum, with their energy
density on average behaving like matter. The initial production of
gravitinos would proceed initially in a mixture of radiation and
matter, but would be later diluted by the entropy produced by the
decay of the singlets. 

\section{Conclusion}
\label{sect5}

The key result of this paper is that the SUSY hybrid model,
in particular Eq. (\ref{superpot2}), has regimes of warm inflation. 
Up to now, it has been assumed that this model in all
parameter regimes  has only cold inflationary dynamics.
However, Fig. (\ref{plot1}) firmly dispels this belief, as it shows that
the parameter regime divides into regions of both warm and
cold inflation.  In light of this finding, the
scalar spectral index, its running, and the tensor-scalar
ratio have been computed in the entire parameter space of
these models.  We find a clean prediction for strong dissipative
warm inflation with $n_S-1=0.98$ and a tensor-scalar ratio
that is effectively zero.  As shown in Fig. (\ref{plot4}), this prediction is
very clearly separated from the cold results,
which up to now have been the expected predictions from these models.
Also these predictions for strong dissipative warm inflation are
clearly separated from those of weak dissipative warm inflation.

Theoretically the effects of dissipation in these models also present 
distinctive features.  In particular, in the
strong dissipative regime there is no $\eta$-problem.
Moreover, even for large coupling $\kappa \sim 1$, the
inflaton field amplitude is well below the Planck scale.  A consequence
of these features is that SUGRA corrections are insignificant.
One interpretation of this would be that a much richer variety
of supergravity extensions of the basic model 
Eqs. (\ref{superpot}) and (\ref{superpot2})
are permissible in comparison to the cold inflation case. 
This could have important model building applications,
especially when identifying viable inflation models in
low-energy limits of string theory.

The main purpose of this paper was to highlight the dissipative
dynamics inherent in this very popular SUSY hybrid model and
then to outline the variety of new features this implies.
There remains a great deal about our results that must be
studied in further detail in future work.  
For example, as shown in \cite{hm} the process of
radiation production will also induce small temperature
dependent corrections to the inflaton effective potential.
These effects will alter predictions
for density perturbations.  In fact, as shown in \cite{hmb1}
temperature dependent effects could introduce qualitatively
new features to the scalar power spectrum, such as oscillations.
Thus a more accurate treatment of density perturbations and
a thorough examination of their evolution is important
to consider in future work.  Along similar lines, a deeper
issue is that of thermalization.  In the analysis in this paper,
we followed the results in \cite{br4,br}, which treats thermalization
based on some simple criteria.  As stated in those works, a
proper dynamical treatment of thermalization is still needed,
and the consequences of such work could make significant changes
in regards the predictions for density perturbations in some
parameter regimes.  However this paper has outlined
the basic result that there are vast parameter regimes in this
model, in which there is particle production during inflation,
and thus in these regimes the statistical state of
the system is substantially altered from the ground state.

For the parameters values of the strong dissipative regime,
reheating may start with a fraction of the vacuum energy
already converted into background radiation.  The decay products of the
inflaton acquire plasma (temperature dependent) masses wich will 
affect the reheating process \cite{riotto}, kinematically blocking the
inflaton decay until the temperature falls below the inflaton mass.   
Processes involving different particle production (thermal or
out-of-equilibrium) mechanisms during reheating should then be reexamined,
such as production of RH (s)neutrinos and leptogenesis. 
In any case, reheating is completed in the warm inflation scenario
through new decay channels different from those active during
inflation. Inflation per se does not force the couplings
entering in Eq. (\ref{superpot2}) to be small as we have seen, and neither
does reheating and the $T_{RH}$ constraint. 

One general result that can be taken away from this paper is that warm
inflation regimes can be expected in SUSY inflationary models. There are
many other models aside from the one studied in this paper in which we
expect to find warm inflation regimes. An interesting case, that is
worth mentioning are SUSY models which lead to monomial inflaton
potentials. One important example of such a model is the
next-to-minimal supersymmetric standard model (NMSSM) \cite{nmssm}, which has
the superpotential  
\be
W=\lambda \Phi H_u H_d - \kappa \Phi^3 + h_t Q_3 U^c_3 H_u + \cdots \,,
\ee 
where $\Phi$ is a singlet superfield, $H_u$, $H_d$ are the Higgs
doblets, $Q_3$ the third generation left-handed quarks and  $U^c_3$
the corresponding right handed up quark, and $h_t$ the top Yukawa coupling. 
Identifying the singlet field $\Phi$ with the inflaton, leads to an
inflaton potential $\sim \kappa^2 \phi^4$. In standard inflation, such
a possibility for the NMSSM has not been of great interest, since it
is well known for such a chaotic inflaton potential that the amplitude
of the inflaton is greater than the Planck scale, $\langle \phi
\rangle > m_P$, thus leading to the problem of an infinite number of
unsuppressed higher dimensional contributions entering the
potential. However in warm inflation, it is known \cite{br} that
monomial potentials like this one yield observationally consistent
warm inflation for field amplitudes {\it below} the Planck scale 
$\langle \phi \rangle < m_P$. Thus, in warm inflation such potentials
have no trouble with higher dimensional contributions, and so are
completely consistent. As such, this fact implies that NMSSM is a model,
with no further modifications, that can support inflation. This is one
of the simplest and may even be the minimal model that is consistent
with the Standard Model and yields inflation. In future work we plan
to do a detailed analysis of warm inflation in the NMSSM. 

\begin{acknowledgments}
The authors thank Q. Shafi for valuable discussions.
AB was funded by the United Kingdom Particle Physics and
Astronomy Research Council (PPARC).
\end{acknowledgments}


\begin{thebibliography}{99}

\bibitem{kk} G. L. Kane and S. F. King,
New Jour. Phys. {\bf 3} (2001) 21. 

\bibitem{shafi} V. N. Senoguz and Q. Shafi, Phys. Lett. B {\bf 567}
(2003) 79.

\bibitem{bck} M. Bastero-Gil, V. Di Clemente and S. F. King,
hep-ph/0408336.

\bibitem{mohapatra} D. Kazanas, R. N. Mohapatra, S. Nasri and
  V. L. Teplitz, hep-ph/0403291. 

\bibitem{jkls}
R. Jeanerot, S. Khalil, G. Lazarides and Q. Shafi, JHEP {\bf 10} (2000) 12.

\bibitem{br4} A. Berera and R. O. Ramos, hep-ph/0406339.

\bibitem{hm} L. M. H. Hall and I. G. Moss, hep-ph/0408323.

\bibitem{br} A. Berera and R. O. Ramos,
Phys. Rev. D{\bf 63} (2001) 103509;
Phys. Lett. B{\bf 567} (2003) 294;
hep-ph/0308211.

\bibitem{wi} A. Berera,  Phys. Rev. Lett. {\bf 75} (1995) 3218;
Phys. Rev. D{\bf 54} (1996) 2519;
Phys.\ Rev.\  D{\bf 55} (1997) 3346.

\bibitem{oldi} A. H. Guth, Phys. Rev {\bf D23} (1981) 347;
K. Sato, Phys. Lett. B{\bf 99} (1981) 66.
                                                                               
\bibitem{ni} A. Albrecht and P. J. Steinhardt, Phys. Rev. Lett.
{\bf 48} (1982) 1220; A. Linde, Phys. Lett. {\bf 108B} (1982) 389.

\bibitem{ci} A. Linde, Phys. Lett. {\bf 129B} (1983) 177.


\bibitem{copeland} E. J. Copeland, A. R. Liddle, D. H. Lyth,
  E. D. Stewart and D. Wands, Phys. Rev. {\bf D49} (1994) 6410.         

\bibitem{dvali} G. Dvali, Q. Shafi and R. Schaefer,
  Phys. Rev. Lett {\bf 73} (1994) 1886.        

\bibitem{lazarides} G. Lazarides, R. K. Schaefer and Q. Shafi,
  Phys. Rev. {\bf D56} (1997) 1324. 

\bibitem{etaproblem} M. Dine, L. Randall and S. Thomas,
  Phys. Rev. Lett. {\bf 75} (1995) 398. 


\bibitem{linderiotto} A. linde and A. Riotto, Phys. Rev. {\bf D56}
  (1997) R1841. 

\bibitem{kawasaki} M. Kawasaki, M. yamaguchi and J. Yokoyama,
  hep-ph/0304161. 

                
\bibitem{COBE} G. F. Smoot et al., Astrophys. J. Lett. {\bf 396}
  (1996) L1; C. L. Bennet et al.,   Astrophys. J. Lett. {\bf 464}
  (1996) 1. 

\bibitem{WMAP} WMAP collab.: D. N. Spergel et al., astro-ph/0302209;
G. Hinshaw et al., astro-ph/0302217;   H. V. Peiris et al., astro-ph/0302225.

\bibitem{panag} C. Panagiotakopoulos, Phys. Rev. {\bf D55}
  (1997) R7335; W. Buchmuller, L. Covi and D. Delepine,
  Phys. Lett. {\bf B491} (2000) 183. 

\bibitem{lyman} M. Viel, J. Weller and M. G. Hoehnelt,
  astro-ph/0407294. 

\bibitem{seljak} U. Seljak et al., astro-ph/0407372.


\bibitem{Guth:ec}
A.~H.~Guth and S.~Y.~Pi,
Phys.\ Rev.\ Lett.\  {\bf 49} (1982) 1110.


\bibitem{Moss:wn}
I.~G.~Moss,
Phys.\ Lett.\ B {\bf 154}, (1985) 120.

\bibitem{Berera:1995wh}
A.~Berera and L.~Z.~Fang,
Phys.\ Rev.\ Lett.\  {\bf 74} (1995) 1912.


\bibitem{Berera:1999ws}
A.~Berera,
Nucl.\ Phys.\ B {\bf 585}, (2000) 666.


\bibitem{arjunspectrum} A. N. Taylor and A. Berera, Phys. Rev. {\bf
D62} (2000) 083517.

\bibitem{hmb1}  L. M. H. Hall, I. G. Moss and A. Berera,
Phys. Rev. D{\bf 69} (2004) 083525.

\bibitem{warmspectrum} W. Lee and L.-Z. Fang, Phys. Rev. {\bf
D59} (1999) 083503; H. P. de Oliveira and S. E. Joras, Phys. Rev. {\bf
D64} (2001) 063513; J. chan Hwang and H. Noh, Class. Quantum
Grav. {\bf 19} (2002) 527  


\bibitem{warmrunning}  L. M. H. Hall, I. G. Moss and A. Berera,
Phys. Lett. B{\bf 589} (2004) 1.

\bibitem{tensor} C. M. Hirata and U. Seljak, Phys. Rev. {\bf
D68} (2003) 083002; 
  U. Seljak and C. M. Hirata, Phys. Rev. {\bf D69} (2004) 043005; 
S. Mollerach,  D. Harari and S. Matarrese, Phys. Rev. {\bf D69} (2004) 063002. 


\bibitem{gravitino} M. Yu. Khlopov and A. D. Linde, Phys. Lett. {\bf
  B138} (1984) 265; J. Ellis, J. E. Kim and D. Nanopoulos,
  Phys. Lett. {\bf B145} (1984) 181. 

\bibitem{gravitino2} For a review, see S. Sarkar,
  Rept. Progr. Phys. {\bf 59} (1995) 1493. 

\bibitem{gravitino3} M. Kawasaki, K. Kohri and T. Moroi,
  astro-ph/0402490. 

\bibitem{kohri} K. Kohri, M. Yamaguch
i and J. Yokoyama, hep-ph/0403043.

\bibitem{lazalepto} G. Lazarides and N. D. Vlachos, Phys. Lett.{\bf
  B459} (1999) 482.  

\bibitem{shafilepto} 
V. N. Senoguz and Q. Shafi, Phys. Lett.{\bf B582} (2004) 6. 

\bibitem{turner} R. J. Scherrer and M. S. Turner, Phys. Rev. {\bf D31}
  (1985) 681. 

\bibitem{riotto} E. W. Kolb, A. Notari and A. Riotto, hep-ph/0307241. 

\bibitem{nmssm}
P. Fayet, Nucl. Phys. {\bf B90} (1975) 104; 
H.-P. Nilles, M. Srednicki and D. Wyler, Phys. Lett. {\bf B120}
(1983) 346;
J.-P. Derendinger and C. A. Savoy, Nucl. Phys. {\bf B237} (1984)
307; 
J. Ellis, J. F. Gunion, H. E. Haber, L. Roszkowski, F. Zwirner,
 Phys. Rev. {\bf D39} (1989) 844; 
L. Durand and J. L. Lopez,  Phys. Lett. {\bf B217} (1989) 463; 
M. Drees,  Int. J. Mod. Phys.  {\bf A4} (1989) 3635;
U. Ellwanger, M. Rausch de Traubenberg, C. A. Savoy,
 Phys. Lett. {\bf B315} (1993) 331 and
 Z. Phys. {\bf C67} (1995) 665;
T. Elliott, S. F. King, P. L. White, 
 Phys. Lett. {\bf B351} (1995) 213.


                                                       
\end{thebibliography}
\end{document}